\documentstyle[twoside,fleqn,epsfig,espcrc2]{article}%

\begin{document}
\renewcommand{\refname}{\normalsize\bf References}
\def\no{\nonumber}
\def\qu{\quad}
\def\qb{\bar{q}}
\def\qbm{\bar{\mbox{q}}}
\def\la{\langle}
\def\ra{\rangle}
\newcommand{\gm}{\gamma}
\newcommand{\be}{\begin{eqnarray}}
\newcommand{\ee}{\end{eqnarray}}
\renewcommand{\th}{\theta}
\newcommand{\Sg}{\Sigma}
\newcommand{\dl}{\delta}
\newcommand{\SSg}{\tilde{\Sigma}}
\newcommand{\eq}{\begin{equation}}
\newcommand{\eqx}{\end{equation}}
\newcommand{\eqn}{\begin{eqnarray}}
\newcommand{\eqnx}{\end{eqnarray}}
\newcommand{\ben}{\begin{eqnaray}}
\newcommand{\een}{\end{eqnarray}}
\newcommand{\f}[2]{\frac{#1}{#2}}
\newcommand{\GG}{{\cal G}}
\renewcommand{\AA}{{\cal A}}
\newcommand{\GR}{G(z)}
 \newcommand{\MM}{{\cal M}}
\newcommand{\BB}{{\cal B}}
\newcommand{\ZZ}{{\cal Z}}
\newcommand{\DD}{{\cal D}}
\newcommand{\HH}{{\cal H}}
\newcommand{\RR}{{\cal R}}
\newcommand{\GT}{{\cal G}_1 \otimes {\cal G}_2^T}
\newcommand{\GGb}{\bar{{\cal G}}^T}
\newcommand{\Du}{{\cal D}_1}
\newcommand{\Dl}{{\cal D}_2}
\newcommand{\zb}{\bar{z}}
\newcommand{\trqq}{\tr_{q\bar{q}}}
\newcommand{\arr}[4]{
\left(\begin{array}{cc}
#1&#2\\
#3&#4
\end{array}\right)
}
\newcommand{\arrd}[3]{
\left(\begin{array}{ccc}
#1&0&0\\
0&#2&0\\
0&0&#3
\end{array}\right)
}
\newcommand{\tr}{\mbox{\rm tr}\,}
\newcommand{\One}{\mbox{\bf 1}}
\newcommand{\pauli}{\sg_2}
\newcommand{\corr}[1]{\la{#1}\rangle}
\newcommand{\br}[1]{\overline{#1}}
\newcommand{\phib}{\br{\phi}}
\newcommand{\psib}{\br{\psi}}
\newcommand{\lm}{\lambda}
\newcommand{\ksi}{\xi}

\newcommand{\Gb}{\br{G}}
\newcommand{\Vb}{\br{V}}
\newcommand{\Gm}{G_{q\br{q}}}
\newcommand{\Vm}{V_{q\br{q}}}

\newcommand{\ggd}[2]{\GG_{#1}\otimes\GG^T_{#2}\Gamma}
\newcommand{\noi}{\noindent}

\newcommand{\cor}[1]{\left\langle{#1}\right\rangle}

%\title{GREEN'S FUNCTIONS IN  NON-HERMITIAN RANDOM MATRIX MODELS $^*$}
\title{Green's Functions in  Non-hermitian Random Matrix
Models\thanks{Based on invited talks  presented by RAJ at the
Marseille CIRM workshop on ``Free probability and applications'' January
1998 and by MAN at the Max Planck Institute workshop on ``Dynamics of
Complex Systems'' May 1999, Dresden, Germany.}
}

\author{%
	Romuald A. Janik%
	\address{Service de Physique Th\'{e}orique,
	CEA Saclay, F-91191 Gif-Sur-Yvette, France \\[2mm]
	$^b$Marian Smoluchowski Institute  of Physics, Jagellonian
	University, 30-059 Krakow, Poland}$^{,b}$,\,%
	\refstepcounter{address}%
	 Maciej A.  Nowak$^{b,}$%
	\address{GSI, Planckstr. 1, D-64291 Darmstadt, Germany \& \\
        \hspace*{1mm} School of Physics, Korea Institute for Advanced Study,
         Seoul 130-012, Korea}
	G\'{a}bor Papp%
	\address{CNR Department of Physics, KSU, Kent, Ohio 44242, USA \& \\
	\hspace*{1mm} HAS Research Group for Theoretical Physics,
	E\"{o}tv\"{o}s University, Budapest, Hungary},\,and
	Ismail Zahed%
	\address{Department of Physics and Astronomy, SUNY, Stony Brook, New
	York 11794, USA}
}
%\date{\today} \maketitle

\begin{abstract}
\hrule
\mbox{}\\[-0.2cm]

\noindent{\bf Abstract}\\

We review some recent techniques for dealing with
non-hermitian random matrix models based on generalized Green's
functions. We introduce the diagrammatic methods in the
hermitian case and generalize them to the non-hermitian
case. The results are illustrated in terms of the eigenvalue
distribution, eigenvector statistics and addition laws.\\[0.2cm]
%{\em PACS}: 05.30.-d,05.40.+j,03.65.Nk, 24.60-k\\[0.1cm]
\hrule
\end{abstract}
%\pacs{PACS numbers: 05.30.-d,05.40.+j,03.65.Nk, 24.60-k}
%\pacs{}
\maketitle

\section{Introduction}

\noindent
Non-hermitian random matrix models {\small (NHRMM)}
permeate a number of
interesting quantum problems~\cite{SOMMERSREV,UPDATENH}, such as: open
chaotic scattering, dissipative quantum maps, neural networks, quantum
dots, non-hermitian localization, directed chaos, turbulence, QCD.

In this short review, based mainly on our recent works,
we go over some novel techniques~\cite{NONHER,DIAG} 
for analyzing a large class of non-hermitian
matrix models with unitary randomness (complex random numbers).

As a main tool for investigating the properties of
NHRMM we choose {\em generalized,
matrix-valued} Green's function.
We start by presenting standard arguments from Green's function theory
for hermitian
random matrix models. Then, using this construction as a reference frame,
we generalize the idea to the case of NHRMM.
Finally,  we unravel several non-trivial properties of generalized  Green's
functions, in particular the link with eigenvector statistics and
non-hermitian addition laws.

\section{Hermitian  RMM and diagrammatic expansion}

The fundamental problem in random matrix theories is to find the
distribution of eigenvalues $\lambda_i $,
in the large N (size of the matrix $H$)
limit, i.e
\be
\rho(\lambda)= \frac{1}{N} \left\langle \sum _{i=1}^N \delta (\lambda -
\lambda_i)
\right\rangle
\label{spect-h}
\ee
where averaging $\left< \ldots \right>$
is done over the ensemble of $N \times N$ random hermitian
matrices
generated with probability
\be
P(H)=e^{-N {\rm Tr} V(H)} .
\label{probab}
\ee
where $V$ is usually a  polynomial in $H$.

It is usually convenient to introduce the Green's function
\be
G(z)=\frac{1}{N} \left\langle {\rm Tr}\, \frac{1}{z-H}\right\rangle \,.
\label{green}
\ee
The eigenvalue distribution is easy to reconstruct from the
discontinuities
of the Green's function, using the relation
\be
\frac{1}{\lambda + i \epsilon} = {\bf P} \frac{1}{\lambda}
-i\pi \delta(\lambda) \,.
\label{dist}
\ee
Indeed,
\eqn
 -\frac{1}{\pi} {\rm Im}\,\lim_{\epsilon \rightarrow 0}
G(\lambda +i\epsilon)= \frac{1}{N} \left\langle {\rm Tr}\,
\delta(\lambda-H)\right\rangle
= &&\nonumber\\
= \frac{1}{N} \left\langle \sum _{i=1}^N \delta (\lambda -
\lambda_i)
\right\rangle= \rho(\lambda) \,. &&
\label{recon}
\eqnx
This can be also recast in the form
\eq
\label{e.dzbar}
\f{1}{\pi}\partial_{\zb} G(z)=\rho(z)
\eqx
where the derivatives are understood in the `distributional' sense.

There are several ways of calculating Green's functions for HRMM.
A number of methods are based on first diagonalizing the random
matrices $H\rightarrow U\Lambda U^\dagger$, then integrating over $U$
using unitary invariance and finally analyzing the resulting integral
over the eigenvalues. We will employ here the diagrammatic method,
based on analogies of RMM to 0+0 dimensional field theory (see
e.g. \cite{hermdiag}). Its
advantage is that it also works when the unitary symmetry is
broken. This happens often in applications
e.g. when the matrices have some block structure or are a sum of a
random and a fixed matrix. We would like to emphasize that despite
being apparently a perturbative method, the structure of the planar
graphs that contribute allow for a resummation of the whole
perturbative series and give the full {\em exact} result in the planar
limit.

A starting point of the analysis is the expression allowing for the
reconstruction of the Green's function from all the moments
$\left\langle {\rm Tr} H^n \right\rangle$,
\eqn
\label{e.series}
G(z) \hspace*{-2.8mm}&=& \hspace*{-3mm}
	\frac{1}{N} \left\langle {\rm Tr}\,\frac{1}{z\!-\!H}\right\rangle
= \frac{1}{N} \sum_n  \frac{1}{z^{n+1}} \left\langle {\rm Tr} H^n
	\right\rangle \nonumber\\
\hspace*{-2.8mm}&=& \hspace*{-3mm}
\frac{1}{N}\!\left\langle\!\! {\rm Tr}\!\left[  \frac{1}{z} +\!%
\frac{1}{z} H \frac{1}{z} +\!\frac{1}{z} H \frac{1}{z} H \frac{1}{z}
 +\!\cdots \right]\!\right\rangle .
\eqnx
We will use the diagrammatic method to evaluate efficiently the sum of
the moments on the right hand side. We note, however, that
in general this series expansion is convergent only in a neighborhood
of $z=\infty$. So the results of the diagrammatic calculation apply
directly only there. For hermitian random matrices this does not pose
a problem. Since the eigenvalues lie
only on some intervals on the real axis, by (\ref{e.dzbar})
the Green's function is a {\em holomorphic function} of $z$
on the complex plane except for the cuts on the real axis.
Therefore we can reconstruct the Green's function everywhere by analytical
continuation. This last step is in fact trivial to perform for
specific ensembles.

For illustration, we consider now the well-known case of a
random hermitian ensemble with Gaussian distribution.
We introduce a  generating function with a matrix source $J$:
\be
Z(J)= \int   dH
 e^{-\frac{N}{2} {\rm
Tr} H^2 + {\rm Tr} H \cdot  J} \,.
\label{partition}
\ee
All the moments follow from $Z(J)$,
\be
\left< {\rm Tr} H^n \right> = \frac{1}{Z(0)} {\rm Tr}\left(
\frac{\partial }{\partial J} \right)^n %\left.
Z(J)\Big|_{J=0}
\ee
and are straightforward to calculate, since in our case
the partition function reads $Z(J)= \exp {\frac{1}{2N} {\rm Tr} J^2}.$

However, instead of calculating all the moments individually alike by  the
Wick theorem,
we could draw them using ``Feynman rules'' derived from our generating
function, and perform a resummation of {\em all} relevant graphs.
The propagator reads
%\eqn
%\la H^a_b H^c_d\ra &=& \frac{\partial }{\partial J^b_a}
%\frac{\partial }{\partial J^d_c}  Z(J)
%=\frac{\partial }{\partial J^b_a}
%\frac{1}{N} J^c_d  Z(J) = \nonumber\\
%&=&\frac{1}{N} \delta^c_b \delta ^a_d \,.
%\label{prop}
%\eqnx
\be
\la H^a_b H^c_d\ra \!=\frac{\partial^2 Z(J)}{\partial J^b_a \partial J^d_c}
=\!\frac1{N} \frac{\partial J^c_d  Z(J)}{\partial J^b_a}
=\!\frac{1}{N} \delta^c_b \delta ^a_d \,.
\label{prop}
\ee
The $1/z$ in (\ref{e.series}) is represented by a horizontal straight line.
We depict the ``Feynman'' rules in Fig.~\ref{fig.rules}.

The diagrammatic expansion of Green's function is visualized
in Fig.~\ref{fig.rainbow}.
Each ``propagator'' brings a factor of $1/N$, and each loop a
factor of $N$. Therefore only planar graphs survive the large $N$ limit.
Introducing the self-energy $\Sigma$ comprising the sum of all
one-particle  irreducible graphs (rainbow-like), the Green's function reads
\eq
G(z)=\f{1}{N}{\rm Tr}\, \f{1}{z-\Sigma(z)} = \frac{1}{z-\Sigma(z)}\,.
\label{e2}
\eqx

In the large $N$ limit the equation for the self energy $\Sigma$,
follows from
resumming the rainbow-like diagrams of Fig.~\ref{fig.rainbow}. The
resulting equation (``Schwinger-Dyson'' equation
of Fig.~\ref{fig.pastur}) encodes pictorially the structure of these
graphs and reads
\eq
\Sigma=G . %=\frac{1}{N} {\rm Tr} \frac{1}{z-\Sigma} .
\label{SD}
\eqx
Equations (\ref{e2}) and (\ref{SD}) give immediately
$G(z-G)=1$,  so the normalizable solution for the Green's function reads
\be
G(z)=\frac{1}{2}(z-\sqrt{z^2-4})
\label{semi1}
\ee
which, via the discontinuity (cut)  leads to Wigner's semicircle \cite{Wigsemi}
for the distribution of the eigenvalues for hermitian random matrices
\be
\rho(\lambda)=\frac{1}{2\pi}\sqrt{4-\lambda^2}.
\ee

\section{Non-hermitian diagrammatics}

Before we present a generalization of the preceeding methods to
non-hermitian random matrices it is instructive to understand what
goes wrong when we apply the standard series expansion
(\ref{e.series}) to the simplest non-hermitian ensemble --- the
Ginibre-Girko one \cite{GinGir,GIRKO}, with non-hermitian matrices
$\MM$, and measure
\be
P(\MM)= e^{-N {\rm Tr} \MM \MM^{\dagger}} \,.
\ee
It is easy to see that all moments vanish $\cor{\tr \MM^n}=0$, for $n>0$
so the expansion (\ref{e.series}) gives the Green's function to be
$G(z)=1/z$ (diagrammatically this follows from the fact that the
propagator $\cor{\MM^a_b \MM^c_d}$ vanishes and hence the self-energy
$\Sigma=0$). The true answer is, however, different. Only for $|z|>1$
one has indeed $G(z)=1/z$. For $|z|<1$ the Green's function is
nonholomorphic and equals $G(z)=\zb$. The reason for the failure is
that in the expression (\ref{green}),
configuration by configuration,
the resolvent displays poles that are scattered
in the complex $z$-plane. In the large $N$ limit, the poles
accumulate in general on finite surfaces,
over which the resolvent is no longer
holomorphic. Therefore we cannot analytically continue to the most
interesting region where
$\partial G/\partial\zb\neq 0$ on the
nonholomorphic surface, with a finite eigenvalue distribution.

This points the way that a successful generalization should work
directly for small $z$ and that some regularization is
necessary. Moreover we would not like to loose the calculational
flexibility of diagrammatic calculations and express the quantities in
terms of some (generalized) moments.

Exploiting the analogy to two-dimensional electrostatics, the following
method~\cite{GIRKO,SOMMERS,HAAKE} has been proposed.
Let us define the ``electrostatic potential''
\be
F=\frac{1}{N} {\rm Tr } \ln [(z-\MM)(\bar{z}-\MM^{\dagger}) +
	\epsilon^2] \,.
\label{els}
\ee
Then
\eqn
&& \hspace*{-5mm}
\lim_{\epsilon \rightarrow 0} \frac{\partial^2 F(z,\bar{z})}%
	{\partial z\partial \bar{z}}
% \frac{\partial}{\partial \bar{z}} F(z,\bar{z}, \epsilon)
=\lim_{\epsilon \rightarrow 0}
\frac{1}{N}\left\langle\!{\rm Tr} \frac{\epsilon^2}{(|z\!-\!\MM|^2
+\!\epsilon^2)^2}\!\right\rangle
 = \nonumber\\
%\frac{\pi}{N} \left\langle {\rm Tr}\ \delta^{(2)}(z\!-\!\MM)
%\right\rangle=
&&=    \frac{\pi}{N} \left\langle \sum_i \delta^{(2)}(z\!-\!\lambda_i)
	\right\rangle
\equiv
\pi \rho(x,y)
\label{spec}
\eqnx
represents Gauss law,
where $z=x+iy$. The last equality involves diagonalizing the
non-hermitian matrix $\MM$, by a linear ({\em non-unitary})
transformation $\MM=L\Lambda L^{-1}$. $F$ can then be expressed in
terms of the eigenvalues $\lambda_i$, and the matrix $V=(L^\dagger
L)^{-1}$. A short calculation gives finally the required equality to the
complex Dirac delta $\dl(z-\lm_i)$.

% and we used the representation of complex
%Dirac delta $\pi  \delta^{(2)}(z)=\lim_{\epsilon \rightarrow 0}
%\epsilon^2/(\epsilon^2 +|z|^2)^2$ and $\MM$ is a
%nonhermitian random matrix.\\

In the spirit of the electrostatic analogy we could define
the Green's function $G(z,\bar{z})$, as an ``electric field''
\be
G%(z, \bar{z})\!
\equiv\!
 \frac{\partial F}{\partial \bar{z}}\!=\!
\frac{1}{N}\!\lim_{\epsilon \rightarrow 0}\!
\left\langle\!{\rm Tr}\!
\frac{\bar{z}-\MM^{\dagger}}{(\bar{z}\!-\!\MM^{\dagger})(z\!-\!\MM)
+\!\epsilon^2)}\!\right\rangle .\!
\label{GG}
\ee
However, instead of working {\it ab initio} with such quantity, and in
view of applying diagrammatic and free-random variable methods
it is much more convenient to proceed differently as we now discuss.

Following~\cite{NONHER,DIAG} we define the matrix-valued resolvent
through\footnote{A slightly different realization was later proposed
in \cite{ZEENEW2} and \cite{Wang}.}
\eqn
\hat{{\cal {G}}}
%={\rm Tr}_{B}\arr{{\cal G}_{qq}}{{\cal G}_{q\overline{q}}}{{\cal G}_{\overline{q}q}}
%{{\cal G}_{\overline{q}\overline{q}}}
&=&\frac{1}{N} \left\langle \rm{Tr_B}
\setlength\arraycolsep{0pt}
\arr{z-\MM}{i \epsilon}{i\epsilon}{\zb
-\MM^{\dagger}}^{-1}\right\rangle \equiv \nonumber\\
&=&
\setlength\arraycolsep{3pt}
\arr{{\GG}_{qq}}{{\GG}_{q\overline{q}}}{{\GG}_{\overline{q}q}}
{{\GG}_{\overline{q}\overline{q}}}
\label{19}
\eqnx
where we introduce the block trace  defined as
\be
{\rm Tr_B}
\setlength\arraycolsep{3pt}
\arr{A}{B}{C}{D}_{2N \times 2N} \hspace*{-3mm}\equiv
\setlength\arraycolsep{3pt}
\arr{
{\rm Tr}\ A}{{\rm Tr}\ B}{{\rm Tr}\ C}{{\rm Tr}\ D}_{2 \times 2}
\hspace*{-5mm}\,.
\ee
Then, by definition, the upper-right component $\GG_{qq}$, is equal to
the Green's function (\ref{GG}).

The block approach has several advantages. Let us define
\be
\label{defzg}
\ZZ=\arr{z}{i\epsilon}{i\epsilon}{\bar{z}} \quad, \quad
\HH=\arr{\MM}{0}{0}{\MM^{\dagger}} \,.
\ee
Then the generalized Green's function is given formally by the same
definition like the usual Green's function $G$,
\be
\GG=\frac{1}{N} \left\langle {\rm Tr_B} \frac{1}{\ZZ-\HH}\right\rangle
	\,.
\label{concise}
\ee
More importantly, also in this case the Green's function is completely
determined by the knowledge of all  matrix-valued moments
\be
\left\langle {\rm Tr_B}\,\,\, \ZZ^{-1} \HH \ZZ^{-1} \HH \ldots \ZZ^{-1}
	\right\rangle \,.
\label{genmom}
\ee
This last observation allows for a diagrammatical interpretation. The
Feynman rules are analogous to the hermitian ones, only now one has to
keep track of the block structure of the matrices.
To demonstrate this, we consider again the case of complex Gaussian
randomness (the Girko-Ginibre ensemble)
\be
P(\MM)= e^{-N {\rm Tr} \MM \MM^{\dagger}} \,.
\ee
In this case the double line propagators are
\eqn
\la \MM^a_b \MM^c_d\ra &=& \la\br{\MM}^a_b \br{\MM}^c_d\ra=0 \quad ,\nonumber\\
\la \MM^a_b \br{\MM}^c_d\ra &=& \la\br{\MM}^a_b \MM^c_d \ra=
	\frac{1}{N}\delta^a_d \delta_c^b \,.
\label{comprop}
\eqnx
As previously, we introduce the self-energy $\tilde{\Sigma}$, (here of course
matrix-valued), in terms of which we get
\be
\GG=(\ZZ-\tilde{\Sigma})^{-1} \,.
\label{gin1}
\ee
The resummation of the rainbow diagrams for $\tilde{\Sigma}$ is more subtle.
Instead of the hermitian equation $\Sigma=G\cdot 0=0$ (due to the
vanishing of the first propagator in (\ref{comprop})), the analogue of
(\ref{SD}) reads now:
\eqn
\tilde{\Sigma} &\equiv&
\setlength\arraycolsep{3pt}
\arr{\Sigma_{qq}}{\Sigma_{q\bar{q}}}{\Sigma_{\bar{q}q}}%
	{\Sigma_{\bar{q}{\bar{q}}}}=
\setlength\arraycolsep{1pt}
\arr{\GG_{qq}\cdot
0}{\GG_{q\bar{q}}}{\GG_{\bar{q}q}}{\GG_{\bar{q}\bar{q}}\cdot 0}=
\nonumber\\
&=&
\setlength\arraycolsep{1pt}
\arr{0}{\GG_{q\bar{q}}}{\GG_{\bar{q}q}}{0}
\,.
\label{gin2}
\eqnx
The diagonal entries are zero, due to the structure of the propagators
(\ref{comprop}). The
two by two  matrix equations~(\ref{gin1}-\ref{gin2}) completely determine the
problem of finding the eigenvalue distribution for complex randomness.
Inserting (\ref{gin2}) into (\ref{gin1}) we get:
\eq\setlength\arraycolsep{3pt}
\arr{{\GG}_{qq}}{{\GG}_{q\overline{q}}}{{\GG}_{\overline{q}q}}
{{\GG}_{\overline{q}\overline{q}}}=
\f{1}{|z|^2-\GG_{q\bar{q}}\GG_{\bar{q}q}} \cdot
\arr{\zb}{{\GG}_{q\overline{q}}}{{\GG}_{\overline{q}q}}{z} .
\label{e.sdgin}
\eqx
Looking at the off-diagonal equation we see that there are two
solutions: first one
with $b^2\equiv \GG_{q\bar{q}}\GG_{\bar{q}q}=0$, second with $b \neq 0$.
The first one is holomorphic, and a straightforward calculation
gives
\be
G(z)=\frac{1}{z} \,.
\ee
The second one is nonholomorhic, and leads, via Gauss law,  to
\be
G(z, \bar{z}) =\bar{z} \, \Longrightarrow \, \rho(x,y)=\frac{1}{\pi}
\frac{\partial}{\partial \bar{z}} G(z,\bar{z}) =\frac{1}{\pi} \,.
\ee
Both solutions match at the boundary $b=0$ , which in this case reads
$z \bar{z}=1$.  In such a simple way we recovered
the results of Ginibre and Girko for the complex non-hermitian ensemble.
The eigenvalues are uniformly distributed on the unit disk $|z|^2<1$.

This simple example illustrates more general properties of matrix valued
function. Each component of the matrix carries important
information about the stochastic properties of the system.
There are always two solutions for $\GG_{qq}$, one holomorphic, another
non-holomorphic. The second one leads, via Gauss law, to the eigenvalue
distribution. The shape of the ``coastline'' bordering
the complex eigenvalues is determined by the
matching conditions for two solutions,
i.e. is determined by the equation
$\GG_{q\bar{q}}\GG_{\bar{q}q}=0$.~\footnote{The shape can be also
inferred from
associated hermitian models using conformal mappings, for details
see \protect\cite{NONHER}.} The product of the off-diagonal
Green's functions  hides another important information about the random
ensemble.
Recently, we have proven~\cite{USEVECT}, that the  correlator
between left and right {\em eigenvectors} (introduced in \cite{CM}) is
given by
\eqn
\label{e.eigv}
O(z)
&\equiv& \cor{ \sum_a (L_a|L_a) (R_a|R_a)\delta (z-\lambda_a)}
=\nonumber\\
&=&-\frac{N}{\pi} \GG_{q\bar{q}} \GG_{\bar{q}q} \,.
\eqnx
For the case of the Girko-Ginibre ensemble the product $\GG_{q\bar{q}}
\GG_{\bar{q}q}$ follows immediately from the matrix
equations~(\ref{gin1}-\ref{gin2}) and reproduces the result \cite{CM}
\eq
O(z)=\f{N}{\pi}\left( 1-|z|^2 \right) \, .
\eqx

\section{Free Random Variables}

One of the powerful calculational features of generalized Green's
functions stems  from direct applicability of  Free Random
Variable (FRV) calculus to this formalism.
Before we demonstrate this, we recall briefly that
the concept of addition law for hermitian ensembles
was introduced in the important work by
Voiculescu~{\cite{VOICULESCU}. In brief, Voiculescu
proposed the additive transformation (R transformation), which linearizes
the convolution of non-commutative matrices, alike the
logarithm of the Fourier transformation  for the convolution
of  arbitrary functions. This method is
an important shortcut to obtain the equations for the Green's
functions
for a sum of matrices, starting from the knowledge of the Green's functions
of individual ensembles of matrices. This formalism was reinterpreted
diagrammatically by
Zee~{\cite{ZEE}, who introduced the concept of Blue's function.

Let us consider the problem of finding the Green's function
of a sum of two  independent (free~{\cite{VOICULESCU})
random matrices $H_1$ and $H_2$,
provided we know the Green's functions of each of them
separately.
First, we note that the 1PI self-energy $\Sigma$ can be always expressed
as a function of $G$ itself and {\em not of z} as usually done in
textbooks. For the Gaussian randomness, $\Sigma (G)=G$ (see (\ref{SD})).
Second, we note that the graphs contributing to the self-energy
$\Sigma_{1+2}(G)$ split into two classes, belonging
to $\Sigma_1(G)$ and $\Sigma_2(G)$, due to the independence of probabilities
$P(H_1)$ and $P(H_2)$ in the large $N$ (planar) limit.
Therefore
\be
 \Sigma_{1+2}(G)= \Sigma_1(G) + \Sigma_2(G) .
\label{sumgg}
\ee
Note that such a formula is not true if the energies are expressed as
functions of $z$. Voiculescu R transformation is nothing but
$R(G)\equiv \Sigma[G]$.
The addition (\ref{sumgg}) reads, for an arbitrary complex $u$,
$R_{1+2}(u)=R_1(u)+R_2(u)$. The R operation forms an abelian group.
The Blue's function, introduced by Zee~{\cite{ZEE}, is simply
\be
B(G)=\Sigma(G)+G^{-1} .
\label{blue}
\ee
Therefore, using the identity $G(z)=1/(z-\Sigma)$, we see that the
Blue's function is the functional inverse of the Green's function
\be
B[G(z)]=z\,.
\label{blueinv}
\ee
In the case of hermitian Gaussian random matrices,
$R(z)=z$, and $B(z)=z+1/z$.

The addition law for Blue's functions reads
\be
B_{1+2}(z)=B_1(z)+B_2(z)-\frac{1}{z} \,.
\label{addblue}
\ee

The algorithm of addition is now surprisingly simple:
Knowing $G_1$ and $G_2$, we find (\ref{blueinv}) $B_1$ and $B_2$.
Then we find the sum $B_{1+2}$ using (\ref{addblue}), and finally,
we get the answer $G_{1+2}$, by reapplying (\ref{blueinv}).
Note that the method treats on equal footing the Gaussian and
non-Gaussian
ensembles, provided that the measures $P_1$ and $P_2$ are independent
(free).

The {\em freeness} of the ensembles, when rephrased in the
diagrammatical
language, means that:\\
(i) Propagators and vertices corresponding to ``Feynman rules'' for
    $H_1$ and $H_2$ ensembles are disjoint, i.e. there are no vertices
which link simultaneously fields of type $H_1$ and $H_2$.

\noindent
(ii) The propagators cannot cross each other.
Large N limit (``asymptotic freeness'') ensures the second condition
 by banning non-planar diagrams.

%*******************************************

The Free Random Variable calculus in the guise of Blue's functions
works extremely well for hermitian random matrix models (see
\cite{ZEE}). However if we were to apply it to the Ginibre-Girko
ensemble obtained as a sum of a gaussian hermitian $\HH_1$ and
antihermitian $i\HH_2$ ensembles, we would run into the same
difficulties which plagued the standard diagrammatic
method. Explicitly the Blue's functions would read\footnote{The factor
$1/2$ reflects just the necessary rescaling of the width of the gaussian.}
\eq
B_{\HH}(z)=\f{1}{2}z+\f{1}{z} \quad , \quad
B_{i\HH}(z)=-\f{1}{2}z+\f{1}{z} \, .
\eqx
Addition law gives then $B_{Ginibre}(z)=1/z$, leading to the result
$G(z)=1/z$ the same as that obtained by blindly applying `hermitian
diagrammatics'.  This shows that one has to generalize the addition
law in order to apply it to NHRMM.

We demonstrated in the first part of these notes,
that by introducing matrix-valued Green's functions we
managed to extend the  diagrammatical analysis to the case of
NHRMM.
Therefore, by imposing the rules (i) and (ii) on {\em non-hermitian}
diagrammatics we can define the freeness condition for non-hermitian
ensembles. This analogous structure allows us to extend now without
much effort the addition formalism to the non-hermitian
case\footnote{Alternatively we could start from the generalized
moments (\ref{genmom}) and apply directly the generalized addition
theorems for ``symmetric moments'' of Voiculescu.}.

The generalized Blue's function~{\cite{NONHER,DIAG}
is now a matrix valued function of
a $2 \times 2$ matrix variable defined by
\be
\BB(\GG)=\ZZ %\equiv \arr{z}{0}{0}{\zb}
\label{e.defbgen}
\ee
where $\ZZ$ and $\GG$ were defined in~(\ref{defzg}) (the $\epsilon$ in
the definition of $\ZZ$ can be here safely set to zero). This is equivalent
to the definition in terms of the self-energy matrix  $\tilde{\Sigma}$,
defined by the general relation~(\ref{gin1}),
\be
\BB (\GG )=\tilde{\Sigma}+\GG^{-1}\,.
\label{genblue}
\ee
The  diagrammatic
reasoning
as before leads to the addition formula for the self-energies and
consequently
for the addition law for generalized Blue's functions~\cite{NONHER,DIAG}
\be
\ZZ  =\BB_1(\GG)+\BB_2(\GG) -\GG^{-1} .
\label{genadd}
\ee

%****************************************************
For illustration we come again to the case of Ginibre ensemble. In
this case the generalized matrix Blue's functions are
\eq
\BB_{\HH}\left[
\setlength\arraycolsep{3pt}
\arr{a}{b}{b}{c} \right] = \f12
\setlength\arraycolsep{3pt}
\arr{a}{b}{b}{c}+ \arr{a}{b}{b}{c}^{-1},
\eqx
\eq
\BB_{i\HH}\left[
\setlength\arraycolsep{3pt}
\arr{a}{b}{b}{c} \right] = \f12
\setlength\arraycolsep{3pt}
\arr{-a}{b}{b}{-c}+ \arr{a}{b}{b}{c}^{-1}\hspace*{-2mm}.
\eqx
The generalized Blue's function of the Ginibre ensemble is thus
\eq
\BB_{Ginibre} \left[
\setlength\arraycolsep{3pt}
\arr{a}{b}{b}{c} \right] =
\setlength\arraycolsep{3pt}
\arr{0}{b}{b}{0}+ \arr{a}{b}{b}{c}^{-1} \hspace*{-3mm} .
\eqx
Solving the defining equation
(\ref{e.defbgen}) is equivalent to (\ref{e.sdgin}) and, as we saw,
gives the full answer.

The power of the addition law for NHRMM stems from the fact that it
treats Gaussian and non-Gaussian randomness on the same footing~\cite{ZEENEW2}.
Also it is a versatile way of obtaining results for non-hermitian
random matrix ensembles starting from some simple building blocks,
without having to go through explicit constructions of (supersymmetric)
sigma-models for the relevant ensembles. In conjunction with
(\ref{e.eigv}) this method provides, in addition, information on the
eigenvector statistics $O(z)$.

\section{Conclusions}

In this short review
we tried to emphasize the versatility of the
generalized Green's function approach and underline
the connections between the hermitian and non-hermitian 
treatment. Several details and practical applications of
the results presented here are included in already published
papers~{\cite{NONHER,DIAG,USMUX}.

%\vskip 0.1cm
\section*{Acknowledgments}

MAN acknowledges interesting discussions with Yan Fyodorov,
Martin Gutzwiller,  Bernard Mehlig, Ingrid Rotter and Karol \.{Z}yczkowski
during the MPI meeting and wishes to thank KIAS for its kind hospitality
during the time this paper has been completed.
RAJ thanks Roland Speicher for discussions during the CIRM workshop.
This work was supported in part  by the US DOE grants DE-FG-88ER40388
and DE-FG02-86ER40251,
by the Polish Government Project (KBN)  grants
2P03B00814, 2P03B01917 and by the Hungarian grant OTKA F026622.

%\setlength{\baselineskip}{15pt}
%\vspace*{-5mm}

\pagebreak
\onecolumn

\section*{Figures}

\begin{figure}[htbp]
%\vspace*{0.98in}
%%%%%%%%%%%%% 1st graph %%%%%%%%%%%%%%%%%%%%%%
\centerline{\psfig{figure=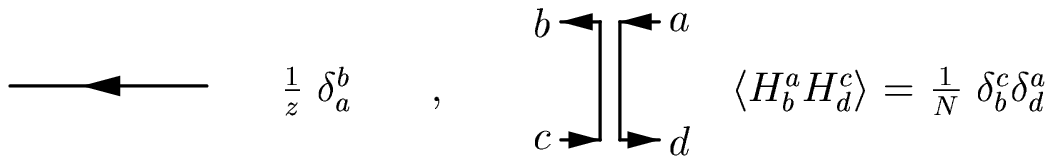,height=15mm}}
%%%%%%%%%%%%% 1st graph %%%%%%%%%%%%%%%%%%%%%%
\caption{Large $N$ ``Feynman rules'' for Gaussian HRMM}
\label{fig.rules}
\end{figure}

\begin{figure}[htbp]
%\vspace*{0.98in}
%%%%%%%%%%%%% 2nd graph %%%%%%%%%%%%%%%%%%%%%%
\centerline{\psfig{figure=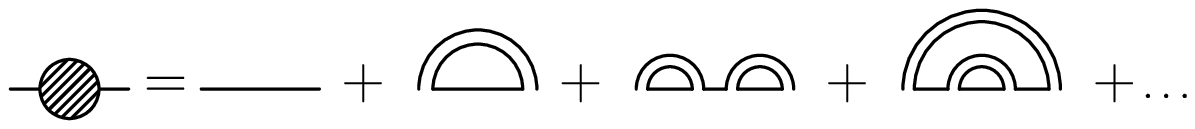,width=115mm}}
%%%%%%%%%%%%% 2nd graph %%%%%%%%%%%%%%%%%%%%%%
\caption{Diagrammatic expansion of Green's function (\protect\ref{green})
for Gaussian ensemble. The $2^{nd}$ and $4^{th}$ graphs are
``rainbow'' graphs contributing to the self-energy $\Sigma$.}
\label{fig.rainbow}
\end{figure}

\begin{figure}[htbp]
%\vspace*{0.98in}
%%%%%%%%%%%%% 3rd graph %%%%%%%%%%%%%%%%%%%%%%
\centerline{\psfig{figure=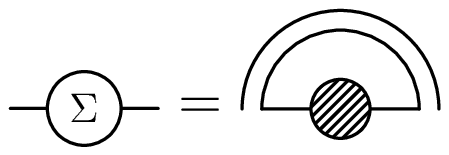,width=40mm}}
%%%%%%%%%%%%% 3rd graph %%%%%%%%%%%%%%%%%%%%%%
\caption{Schwinger-Dyson equation.}
\label{fig.pastur}
\end{figure}

\end{document}